# Quantum light sources with configurable lifetime leveraging parity-time symmetry

Nuo Chen,[1,9] Wen-Xiu Li,[2,3,9] Yun-Ru Fan,[2,3,*] Hang-Hang Li,[1] Hong Zeng,[2,3] Wu-Qiang Chi,[1] Heng Zhou,[4] Hao Li,[5] Li-Xing You,[5] Guang-Can Guo,[2,3,6,7] Qiang Zhou,[2,3,6,7,†] Jing Xu,[1,8,‡] and Xin-Liang Zhang[1,8,§]

[1]School of Optical and Electronic Information & Wuhan National Laboratory for Optoelectronics, Huazhong University of Science and Technology, Wuhan 430074, China

[2]Institute of Fundamental and Frontier Sciences, University of Electronic Science and Technology of China, Chengdu 611731, China

[3]Key Laboratory of Quantum Physics and Photonic Quantum Information, Ministry of Education, University of Electronic Science and Technology of China, Chengdu 611731, China

[4]Key Lab of Optical Fiber Sensing and Communication Networks, School of Information and Communication Engineering, University of Electronic Science and Technology of China, Chengdu 611731, China

[5]National Key Laboratory of Materials for Integrated Circuits, Shanghai Institute of Microsystem and Information Technology, Chinese Academy of Sciences, Shanghai 200050, China

[6]Center for Quantum Internet, Tianfu Jiangxi Laboratory, Chengdu 641419, China

[7]CAS Key Laboratory of Quantum Information, University of Science and Technology of China, Hefei 230026, China

[8]Optics Valley Laboratory, Hubei 430074, China

[9]These authors contributed equally: Nuo Chen, Wen-Xiu Li

[*]Corresponding author: Yun-Ru Fan (yunrufan@uestc.edu.cn)

[†]Corresponding author: Qiang Zhou (zhouqiang@uestc.edu.cn)

[‡]Corresponding author: Jing Xu (jing_xu@hust.edu.cn)

[§]Corresponding author: Xin-Liang Zhang (xlzhang@hust.edu.cn)

**Abstract** Quantum light sources with configurable photon lifetimes are essential for large-scale quantum circuits, enabling applications in programmable quantum computing, various quantum key distribution protocols, and quantum tomography techniques. However, the fundamental trade-off between efficiency and photon lifetime imposes significant challenges on the design of high-performance large configurable lifetime quantum light sources. Here, we report on such chip-scale quantum light sources by harnessing the unique feature of parity-time (PT) symmetry. The core design centers on employing PT-symmetric coupling between two microresonators of distinct circumferences, enabling broad-range and selective tuning of intracavity photon density of states. By controlling the alignment between resonators, we achieved a 38-fold photon lifetime tuning range (4 ~ 158 ps), with the shortest lifetimes near the exceptional points of the PT-symmetric systems. The device generates energy-time entangled photon pairs with $87.1 \pm 1.1\%$ interference visibility and a heralded second-order autocorrelation of $g_h^{(2)}(0) = 0.069 \pm 0.001$. Our work highlights the potential of PT symmetry for advanced quantum applications, including high-speed communication and programmable quantum computing, quantum coherent tomography, and beyond.

## Introduction

On-chip generation of diverse optical quantum states has been widely demonstrated, highlighting its substantial potential for advancing integrated quantum information processing applications [1–8]. As the field advances, the scalable and configurable generation and manipulation of quantum states across various degrees of freedom become increasingly essential [5,6,9–19]. Photon lifetime is one of the critical degrees of freedom in optical quantum states, predominantly determined by the bandgap of a two-level system or the cavity quality factor (Q-factor) of a resonator-based photon emitter [20]. For example, photons with long lifetimes, on the order of several nanoseconds, typically exhibit relatively narrow linewidths, ranging from a few MHz to several hundred MHz. They are indispensable for aligning with the atomic transition linewidths to enable effective photon-atom interactions, crucial for quantum repeater-based long-distance quantum secure communications [21–22]. On the other hand, to maximize the rate of quantum communication systems, the lifetime of single-photons $\tau$ must be compressed, since the communication bandwidth is proportional to $2\pi/\tau$, as governed by Fourier reciprocity [13]. Short-lived photons are also required in quantum optical coherent tomography [24,25]. Furthermore, the ability to configure photon lifetime is vital for a variety of applications. Dynamically tunable photon lifetime schemes have been utilized as quantum random number generators [9]. Sources with configurable photon lifetime are highly promising for frequency-encoded qubits, a key component in scalable quantum information processing technologies [26].

On the integrated photonics platform, research has shown that the control of the photon lifetime can be realized by integrated modulation modules to shape spectra of single-photon pulses [9,12–16]. However, implementing modulation after photon generation complicates the quantum system structure and can introduce extra losses. In recent years, optical microresonators utilizing spontaneous nonlinear effects have been extensively demonstrated as quantum light sources [1–3,27–34]. It has been shown that different photon lifetimes can be achieved by setting distinct coupling gaps between the resonator and the bus waveguide [20]. Nevertheless, modifying the photon lifetime becomes challenging once the quantum light source devices are fabricated. More importantly, the fundamental trade-off between efficiency and bandwidth imposes significant constraints on the design of high-performance configurable photon lifetime quantum light sources. That is, to generate photons with a short lifetime, the duration of light-matter interaction must be reduced (related to an increase of the cavity bandwidth), which in return severely deteriorates the generation efficiency [35,36]. Therefore, it is crucial to find other photon lifetime adjustment schemes that do not sacrifice the efficiency, or the photon generation rate (PGR).

Recent advances in parity-time (PT)-symmetric systems have attracted substantial attention due to their easy implementation yet profound implications in photonics [37–41]. The non-Hermitian property of PT-symmetric systems, especially at the exceptional point (EP) where the eigenvalues and corresponding eigenvectors coalesce, gives rise to a range of unconventional light-matter interaction phenomena [42,43]. These properties enable wide applications in areas such as lasing [44], enhanced sensing [45,46], coherent absorption [47], optical frequency combs [48,49], and beyond. In addition, we have shown in a recent work that by leveraging PT symmetry and breaking the uniformity of resonance linewidth of the coupled system, the trade-off between efficiency and bandwidth has been relieved for classical applications such as nonlinear optical signal processing [50]. This indicated that the trade-off between short photon lifetime and large PGR may be relieved via a similar linewidth-manipulation method. Indeed, integrated resonance-manipulated devices have been studied for several quantum applications, such as spectral-uncorrelated photon-pairs generation [51,52] or high-coincidence efficiency photon-pairs sources [53]. Nevertheless, achieving direct and large-range control of cavity photon lifetime remains elusive.

In this work, we design an effective photon lifetime-configurable quantum light source by leveraging a dual-ring coupled PT-symmetric system. PT symmetry introduces an extra degree of freedom through unbalanced losses to control the decay rates of microresonators, making it naturally suitable for adjusting the photon lifetime. The maximum loss contrast near and away from the EP enables orders of magnitude control of the photon lifetime. Meanwhile, the PGR has maintained a moderately high-value thanks to the undegraded pump enhancement via the linewidth manipulation technique. Experimentally, we demonstrate a 38-fold variation in photon lifetime, from 4.1 to 158.4 ps. The generation of energy-time entangled photon pairs exhibits a typical raw two-photon interference visibility of $87.1 \pm 1.1\%$. A typical heralded second-order autocorrelation $g_h^{(2)}(0)$ of $0.069 \pm 0.001$ is also obtained.

## Results

### System design and principle of operation

We demonstrate the configurable photon lifetime quantum light source using spontaneous four wave-mixing (SFWM) in a dual-ring coupled PT-symmetric system, as sketched in Fig. **1a**. The main ring (red) with a radius $R_1$ is coupled to the auxiliary ring (blue) with a radius $R_2$. The intrinsic decay rates of the main and auxiliary rings are $\gamma_1$ and $\gamma_2$, respectively, and the coupling strength between the two rings is $\kappa$. The auxiliary ring is coupled to the bus waveguide with the coupling decay rate $\gamma_c$, resulting in a total decay rate of the auxiliary ring as $\gamma_1 + \gamma_c$. When $\gamma_c \gg \gamma_{1,2}$, the decay rates of the two coupled rings become severely unbalanced, and a passive PT-symmetric system is thus formed [48]. Given that the radius ratio is set to $R_1 = 2R_2$, the free spectral range (FSR) of the two rings (denoted as $FSR_1$ and $FSR_2$, respectively) differs by a factor of 2, as depicted by the intracavity photon density of state (DOS) in Fig. **1b**. A thermal tuning module is fabricated on the auxiliary ring to adjust the detuning between the two rings. When a voltage is applied to the thermal tuning electrode, the resonance of the auxiliary ring gradually aligns with that of the main ring, consequently reducing the Q-factors of the corresponding resonances of the main ring (see Fig. **1c**(ii) and **1d**). The resonances influenced by this adjustment are designated as the signal and idler resonances, whereas the central resonance, which remains unchanged due to the FSR difference of the two rings, functions as the pump resonance (see Fig. **1d**(iii)). The purpose of this design is twofold. On one hand, the reduced Q-factors of the signal and idler resonances result in a shorter photon lifetime. On the other hand, the unaffected high-Q pump resonance ensures high SFWM efficiency in the quantum light source, thereby maintaining a high PGR. Physically, this phenomenon arises because spontaneous four-wave mixing (SFWM) primarily occurs in the main ring, where the pump light is predominantly confined [50], while the generated photons are emitted through the bus waveguide. The photon lifetime of the emitted photons is largely determined by the coupling decay introduced by the auxiliary ring.

Without loss of generality, we consider in the following the evolution of signal resonance since the signal and idler resonances act similarly. According to temporal coupled-mode theory (TCMT), the non-Hermitian Hamiltonian of the system reads [40,50]

$$\widehat{H} = \begin{bmatrix} \omega_1 - i\frac{\gamma_1}{2} & \kappa \\ \kappa & \omega_2 - i\frac{\gamma_2 + \gamma_c}{2} \end{bmatrix},$$

(1)

where $\omega_1$, $\omega_2$ are the resonant frequencies of the main ring and the auxiliary ring, respectively. $\delta\omega = \omega_2 - \omega_1$ is the detuning between the two rings. Solving the complex eigenfrequencies of Hamiltonian, the eigenfrequencies of the PT-symmetric system can be expressed as [40,50]

$$\omega_{\pm} = \left[\frac{1}{2}(\omega_1 + \delta\omega) - \frac{i}{4}(\gamma_2 + \gamma_c + \gamma_1)\right] \pm \frac{1}{4}\sqrt{16\kappa^2 - (\gamma_2 + \gamma_c - \gamma_1)^2}.$$

(2)

Discussion simplifies when the intrinsic losses of the two rings are identical, i.e., $\gamma_1 = \gamma_2$, which can be approximated in our design and fabricated devices. Under the condition of $\delta\omega = 0$ (i.e., the two rings are aligned), the EP occurs at the degeneracy point of complex eigenfrequencies, i.e., $\kappa = \gamma_c/4$ (for details, see Supplementary information S1: System analysis methods).

Consider a typical situation that commonly occurs in integrated coupled ring systems when the coupling between two rings is strong and ring-wave coupling loss is small, i.e., $\kappa > \gamma_{1,2} \gg \gamma_c$ [9]. The system exhibits a typical mode splitting that corresponds to $\kappa$ at $\delta\omega = 0$, as mapped in Fig. **1c**(i). In contrast, the losses of the coupling decay $\gamma_c$ in our design is much larger than the intrinsic losses $\gamma_{1,2}$. When the coupling coefficient is set to be $\kappa \approx \gamma_c/4$, the system operates near the EP, as mapped in Fig. **1c**(ii). A merged single resonance with a significant low-Q is formed at $\delta\omega = 0$, which is naturally suitable for generating short-lived photons.

The photon lifetime adjustment process is achieved by changing the detuning between two rings using a thermal tuning module, as illustrated in Fig. **1d**. Initially, the auxiliary ring with resonant frequency $\omega_2$ is misaligned with the main ring $\omega_1$. The signal and idler resonances can be regarded as resonances from a single ring under critical coupling with a linewidth of $\Delta\omega = 2\gamma_1$ when detuning is large, e.g., $\delta\omega \approx FSR_1/2$ [50]. Therefore, the signal and idler resonances are high-Q resonances in this case, as shown in Fig. **1d**(i), and the photon lifetime is calculated by [20,54]

$$\tau_{\text{high-Q}} = \frac{1}{\Delta\omega} = \frac{1}{2\gamma_1}.$$

(3)

As the voltage applied to the thermal tuning electrode of the auxiliary ring gradually increases, the resonance of the auxiliary ring approaches that of the main ring (Fig. **1d**(ii)) until they are completely aligned ($\delta\omega = 0$), as shown in Fig. **1d**(iii). In case (iii), the signal and idler resonances operate near the EP, with a linewidth of $2\text{Im}(\omega_+) \approx \gamma_c$, according to Eq. (**2**). At this point, the photon lifetime is given by

$$\tau_{\text{low-Q}} = \frac{1}{\gamma_c}.$$

(4)

It can be derived from Eqs. (**3**) and (**4**) that the maximum achievable contrast of the photon lifetime is $\gamma_c/2\gamma_1$. Thus, increasing the coupling decay rate of the auxiliary and the bus-waveguide or reducing the intrinsic loss of the main ring can enhance the contrast ratio in the tunability of the photon lifetime.

## Device characterization

We then design the device on the low-loss integrated silicon nitride-on-insulator (SNOI) platform to experimentally demonstrate a configurable photon lifetime quantum light source. The chip is fabricated by LIGENTEC (Switzerland) using a standard thick silicon nitride waveguide process. The waveguide dimensions are 1000 nm × 800 nm (width × height), supporting only the fundamental mode. The radius of the main ring is designed as $R_1 = 228\ \mu\text{m}$ (with an FSR close to 100 GHz to match dense wavelength division multiplexer (DWDM) channels), and the radius of the auxiliary ring is $R_2 = 114\ \mu\text{m}$. The transmission spectra from 1534 nm to 1540 nm of the system under different voltages, i.e., 0 V and 6.6 V, are plotted in Fig. **2a** and Fig. **2b**, respectively (for details, see Supplementary information S2: Spectra characterizations). Under a 0 V bias, the two rings are misaligned, and the overall transmission spectrum exhibits a series of near critical-coupled high-Q resonances from the main ring ($FSR_1 = 0.78$ nm). Between each two high-Q resonances, there are auxiliary ring resonances (yellow lines) with a relatively low extinction ratio ($FSR_2 = 1.56$ nm) due to the auxiliary ring being in a state of over-coupling with the bus waveguide. Under a bias of 6.6 V, the resonances of the main ring and the auxiliary ring are aligned, resulting in a distribution of alternating low-Q and high-Q resonances in the transmission spectrum. Resonances that are three $FSR_1$ away from the pump resonance on both sides are selected as the signal and idler resonances. To be concrete, the resonance centered at 1536.9 nm (C51 DWDM channel) is chosen as the pump resonance. The resonances centered at 1534.6 nm (C54) and 1539.3 nm (C48) are chosen as the signal and idler resonances, respectively, as shown by the dashed areas in Fig. **2a** and **2b**.

The Q-factors of all resonances are fitted and marked by the green and orange circles (the right axis). Before and after thermal tuning, the Q-factors of the resonances that coupled to the auxiliary ring exhibit a significant reduction, whereas the unaffected resonances keep a similar high-Q factor under critical coupling (see Supplementary information S3: Critical coupling condition of pump resonance). According to TCMT, the system parameters are extracted by the fittings: $\gamma_1 = \gamma_2 = 3.0$ GHz, $\gamma_c = 146.8$ GHz, $\kappa = 45.5$ GHz. It can be seen that $\kappa = \gamma_c/4$ is approximately satisfied, indicating that the system operates near the EP. Therefore, the expected photon lifetimes before and after the alignment are $\tau_{high-Q} = 167.7$ ps and $\tau_{low-Q} = 6.5$ ps, respectively, according to the relation between decay rate and photon lifetime (Eqs. **(3)** and **(4)**).

## Correlated photon-pairs generation

Next, we characterize the photon lifetime of the correlated photon pairs generated from the lifetime configurable device. Figure **3a** plots the experimental setups for correlated photon-pairs generation. The pump light from a continuous wave (CW) laser is passed through an erbium-doped fiber amplifier (EDFA), a variable optical attenuator (VOA), and a polarization controller (PC), and then into a 90/10 beam splitter (BS). 10% of the pump enters a power meter (PM), to monitor the input power entering the chip. The remaining 90% of the pump is filtered out using a high-isolation (with an isolation of ⩾120 dB) DWDM at the C51 channel and then coupled into the chip. The DWDM is employed here to filter out noise photons at the signal and idler wavelengths of the preceding system. The second DWDM is set after the generation of correlated photon pairs as the pump rejection filter. Figure **3b** gives the test system of correlated photon pairs. A tunable bandwidth filter (TBF) is applied to separate the signal and idler wavelengths, and then send the signal and idler photons to the superconducting nanowire single-photon detectors (SNSPDs, with a detection efficiency of 90% and a dark count rate of about 30 Hz). The output signals of SNSPDs are further connected to a time-to-digital converter (TDC) to record the coincidence events.

The normalized coincidence counts (CCs) of generated photon pairs under the bias voltages of 0V and 6.6 V are recorded in Fig. **4a** and **4b**, respectively. For an idler-start-and-signal-stop coincidence detection, assuming signal and idler photons feature the same lifetime, the normalized CCs are fitted with a double exceptional function, (red and blue lines) and their $1/e$ widths are determined [20,55]. Under the 0 V bias, the signal and idler resonances are in high-Q cases, with $\tau_{1/e} = 239.4$ ps. Under the 6.6 V bias, the Q-factors of signal and idler resonances are decreased, with $\tau_{1/e} = 91.9$ ps. Experimentally, because the $1/e$ width of the CCs diagram includes the time jitters of SNSPDs and TDC, the actual photon lifetime $\tau$ is extracted according to the following equation [9]

$$\tau_{1/e} = \sqrt{2\tau^2 + J_{\text{channel 1}}^2 + J_{\text{channel 2}}^2},$$

(5)

where $J_{\text{channel 1}}$ and $J_{\text{channel 2}}$ are the time jitters of the detection channels (together with a single SNSPD and TDC) of signal and idler, respectively, measured as $J_{\text{channel 1}} = 74.5$ ps and $J_{\text{channel 2}} = 53.5$ ps by the autocorrelation tests (see Supplementary information S4: Experimental details and extraction of photon lifetimes). Thus, the photon lifetimes extracted from Eq. **(5)** under 0 V and 6.6 V bias are 156.4 ps and 4.1 ps, respectively. A 38-fold photon lifetime adjusting contrast is achieved, agreeing with the expected variations from the TCMT discussions given earlier.

Next, Fig. **4c** and **4d** draw the peak values of CCs with respect to the on-chip pump powers, from 0.25 mW to 14.12 mW. The CCs scale with the square of the pump power, aligning with the theoretical predictions for the PGR in SFWM [35]. The error bars are derived from Poissonian photon-counting statistics [28,32]. The corresponding coincidence-to-accidental ratios (CARs) are then recorded in Fig. **4e** and **4f**. Under the 0 V bias, the maximum CAR reaches 31, while under the 6.6 V bias, the CAR reaches 170. The increase in the CCs of photons after lowering the Q-factors of the signal resonance, which is unexpected at first sight, is due to several facts. One is that our particular design enables greatly enhanced mixing processes compared to schemes that lowered cavity linewidth for all resonances, i.e. pump, signal, and idler resonances. Second, the PGR in SFWM is proportional to the integral of the spectral photon DOS [33] of signal and idler resonances, resulting in more photon pairs collected over a wider spectrum. In addition, the device is transitioned from critical-coupling to over-coupling in the latter case, rendering more photon pairs can escape from the cavities [36,53].

**Energy-time entanglement and heralded single photons generation**

We tested the energy-time entanglement of the short-lived correlated photon pairs generated by CW pumping. The experimental setup for two-photon interference using an unbalanced Michelson interferometer (UMI) is shown in Fig. **3c**. Figures **5a** and **5b** give the CCs of constructive and destructive two-photon interference. The single side counts of signal and idler photons are recorded in the upper panel of Fig. **5c**, which are near constants (about 40.1 kHz and 40.0 kHz on average, respectively), indicating that single-photon interference does not occur. The measured interference fringe is further shown in the lower panel of Fig. **5c**. The green circles are the coincidence events in 20 seconds, and the dashed green curve is the fit using a 1000-time Monte Carlo method [28,32]. The interference visibility is calculated as 87.1 ± 1.1% (without subtracting the accidental coincidence counts), which exceeds 70.7%, indicating effective energy-time entanglement violating Bell inequality [56].

Last, the heralded single photons can be obtained based on the photons generated in pairs. The heralded second-order autocorrelation $g_h^{(2)}(\tau)$ is measured by a Hanbury Brown–Twiss (HBT) interferometer [57], as shown in Fig. **3d**. In the experiment, the signal photons are detected directly by SNSPD, while the idler photons are detected with a delay time after passing through the 50:50 beam splitter and the threefold coincidence events are recorded by TDC. The $g_h^{(2)}(\tau)$ of heralded single photons (idler) and their heralding photons (signal) is given in Fig. **6**. The zero-delay value $g_h^{(2)}(0)$ is measured to be 0.069 ± 0.001 with a heralding rate of 256 kHz under an 11 mW on-chip pump, indicating that it can be used as a short-lifetime heralded single-photon source.

# Discussion

We experimentally demonstrated a lifetime configurable quantum light source on a SNOI chip harnessing PT-symmetry. By designing a coupled micro-ring structure with a radius ratio of 2:1 and fabricating a thermal tuning electrode on the auxiliary ring to control the detuning between the rings, the device that operates near or away from the EP can realize efficient lifetime adjustment nearly without sacrificing SFWM efficiency. The tuning range of the photon lifetime that we achieved in this experiment is limited by the relatively low intrinsic Q-factor of the main microresonator. Further reducing the waveguide loss, for example by widening the waveguide of microring or using a low-loss fabricating process, can enhance the intrinsic Q-factor of the microcavity, thereby allowing for a greater range of photon lifetime adjustment (more details refer to Supplementary information S5: Device design). As verified by the experimental results, we emphasize that our design enables independent photon lifetime manipulation of the pump and signal (idler) resonances, enabling the adjustment of photon lifetime while maintaining a high PGR. This is highly beneficial for achieving high-quality photon sources with configurable lifetimes for practical applications. We envision that the lifetime-configurable single-photon source based on PT-symmetric coupled microring systems, due to its unique characteristics, can be applied in integrated quantum optics. This includes applications in high-speed quantum key distribution [58–60], frequency-encoded quantum communication [61], quantum computing [62], and the generation of quantum random numbers [10].

## Data availability

The data that support the plots within this paper and other findings of this study are available at XX.

## Acknowledgment

This work is supported by the National Natural Science Foundation of China (NSFC) (Nos. 62275087, 62475037) and Sichuan Science and Technology Program (Nos. 2023YFSY0058, 2023YFSY0059, 2023YFSY0060, 2024YFHZ0368, 2024YFHZ0370);

## Author contributions

Q.Z. and J.X. conceived the original idea. N.C. and H.-H. L. developed the theory. Y.-R. F., Q.Z. and J.X. supervised the theoretical aspects. N.C., H.-H. L., W.-Q.C. and J. X. conceived the device design. N.C. drafted the chip layout. Heng Z., H. L. and L.-X.Y. provided the experimental setups. N.C., W.-X.L. and Hong Z. performed the experimental measurements. N.C., W.-X.L. and Y.-R. F. completed the experimental data analysis. Y.-R. F., Q.Z. and J.X. supervised the experiments. N.C. and W.-X.L. wrote the manuscript with the help of Y.-R. F., Q.Z. and J.X. G.-C.G., Q.Z., J.X. and X.-L. Z. coordinated and supervised the project. All authors commented on the manuscript.

## Competing interests

The authors declare no competing interest.

## Additional information

**Supplementary information** The online version contains supplementary material available at XX.

# Reference

[1]    Caspani, L. et al. Integrated sources of photon quantum states based on nonlinear optics. *Light: Sci. & Appl.* **6**, e17100 (2017).

[2]    Kues, M. et al. On-chip generation of high-dimensional entangled quantum states and their coherent control. *Nature* **546**, 622–626 (2017).

[3]    Reimer, C. et al. Generation of multiphoton entangled quantum states by means of integrated frequency combs. *Science* **351**, 1176–1180 (2016).

[4]    Cogan, D., Su, Z.-E., Kenneth, O. & Gershoni, D. Deterministic generation of indistinguishable photons in a cluster state. *Nat. Photon.* **17**, 324–329 (2023).

[5]    Clementi, M. et al. Programmable frequency-bin quantum states in a nano-engineered silicon device. *Nat. Commun.* **14**, 176 (2023).

[6]    Mahmudlu, H. et al. Fully on-chip photonic turnkey quantum source for entangled qubit/qudit state generation. *Nat. Photon.* **17**, 518–524 (2023).

[7]    Yang, Z. et al. A squeezed quantum microcomb on a chip. *Nat. Commun.* **12**, 4781 (2021).

[8]    Jia, X. et al. Continuous-variable multipartite entanglement in an integrated microcomb. *Nature* doi: https://doi.org/10.1038/s41586-025-08602-1 (2025).

[9]    Joshi, C. et al. Picosecond-resolution single-photon time lens for temporal mode quantum processing. *Optica* **9**, 364 (2022).

[10]    Okawachi, Y. et al. Dynamic control of photon lifetime for quantum random number generation. *Optica* **8**, 1458 (2021).

[11]    Lukin, D. M, et al. Spectrally reconfigurable quantum emitters enabled by optimized fast modulation. *npj Quantum Inf.* **6**, 80 (2020).

[12]    Kielpinski, D., Corney, J. F. & Wiseman, H. M. Quantum Optical Waveform Conversion. *Phys. Rev. Lett.* **106**, 130501 (2011).

[13]    Moiseev, S. A. & Tittel, W. Temporal compression of quantum-information-carrying photons using a photon-echo quantum memory approach. *Phys. Rev. A* **82**, 012309 (2010).

[14]    Zhu, D. et al. Spectral control of nonclassical light pulses using an integrated thin-film lithium niobate modulator. *Light: Sci. & Appl.* **11**, 327 (2022).

[15] Huang, Y. et al. Frequency-insensitive spatiotemporal shaping of single photon in multiuser quantum network. *npj Quantum Inf.* **9**, 83 (2023).

[16] Karpiński, M, Jachura, Wright, L, J & Smith, B, J. Bandwidth manipulation of quantum light by an electro-optic time lens. *Nat. Photon.* **11**, 53 (2017).

[17] Mittal, S., Orre, V. V., Goldschmidt, E. A. & Hafezi, M. Tunable quantum interference using a topological source of indistinguishable photon pairs. *Nat. Photon.* **15**, 542–548 (2021).

[18] Hua, X. et al. Configurable heralded two-photon Fock-states on a chip. *Opt. Express* **29**, 415 (2021).

[19] Sultanov, V. et al. Tunable entangled photon-pair generation in a liquid crystal. *Nature* **631**, 294–299 (2024).

[20] Ramelow, S. et al. Silicon-Nitride Platform for Narrowband Entangled Photon Generation. *arXiv* Preprint at https://doi.org/10.48550/arXiv.1508.04358 (2015).

[21] Chen, R. et al. Ultralow-Loss Integrated Photonics Enables Bright, Narrowband, Photon-Pair Sources. *Phys. Rev. Lett.* **133**, 083803 (2024).

[22] Ortiz-Ricardo, E. et al. Submegahertz spectral width photon pair source based on fused silica microspheres. *Photon. Res.* **9**, 2237 (2021).

[23] Liu, J., Liu, J., Yu, P. & Zhang, G. Sub-megahertz narrow-band photon pairs at 606 nm for solid-state quantum memories. *APL Photonics* **5**, 066105 (2020).

[24] Arahira, S., Murai, H. & Sasaki, H. Generation of highly stable WDM time-bin entanglement by cascaded sum-frequency generation and spontaneous parametric downconversion in a PPLN waveguide device. *Opt. Express* **24**, 19581 (2016).

[25] Nasr, M. B., Saleh, B. E. A., Sergienko, A. V. & Teich, M. C. Demonstration of Dispersion-Canceled Quantum-Optical Coherence Tomography. *Phys. Rev. Lett.* **91**, 083601 (2003).

[26] Lukens, J. M. & Lougovski, P. Frequency-encoded photonic qubits for scalable quantum information processing. *Optica* **4**, 8 (2017).

[27] Labonté, L. et al. Integrated Photonics for Quantum Communications and Metrology. *PRX Quantum* **5**, 010101 (2024).

[28] Zeng, H. et al. Quantum Light Generation Based on GaN Microring toward Fully On-Chip Source. *Phys. Rev. Lett.* **132**, 133603 (2024).

[29] Ma, Z. et al. Ultrabright Quantum Photon Sources on Chip. *Phys. Rev. Lett.* **125**, 263602 (2020).

[30] Lu, X. et al. Chip-integrated visible–telecom entangled photon pair source for quantum communication. *Nat. Phys.* **15**, 373–381 (2019).

[31] Lu, L. et al. Three-dimensional entanglement on a silicon chip. *npj Quantum Inf.* **6**, 30 (2020).

[32] Fan, Y.-R. et al. Multi-Wavelength Quantum Light Sources on Silicon Nitride Micro-Ring Chip. *Laser & Photonics Rev.* **17**, 2300172 (2023).

[33] Jaramillo-Villegas, J. A. *et al.* Persistent energy–time entanglement covering multiple resonances of an on-chip biphoton frequency comb. *Optica* **4**, 655 (2017).

[34] Steiner, T. J. et al. Ultrabright Entangled-Photon-Pair Generation from an Al Ga As -On-Insulator Microring Resonator. *PRX Quantum* **2**, 010337 (2021).

[35] Guo, K. et al. Generation rate scaling: the quality factor optimization of microring resonators for photon-pair sources. *Photon. Res.* **6**, 587 (2018).

[36] Wu, C. et al. Optimization of quantum light sources and four-wave mixing based on a reconfigurable silicon ring resonator. *Opt. Express* **30**, 9992 (2022).

[37] Bender, C. M. & Boettcher, S. Real Spectra in Non-Hermitian Hamiltonians Having PT Symmetry. *Phys. Rev. Lett.* **80**, 5243–5246 (1998).

[38] Guo, A. et al. Observation of P T -Symmetry Breaking in Complex Optical Potentials. *Phys. Rev. Lett.* **103**, 093902 (2009).

[39] Feng, L., El-Ganainy, R. & Ge, L. Non-Hermitian photonics based on parity–time symmetry. *Nature Photon.* **11**, 752–762 (2017).

[40] Özdemir, Ş. K., Rotter, S., Nori, F. & Yang, L. Parity–time symmetry and exceptional points in photonics. *Nat. Mater.* **18**, 783–798 (2019).

[41] Peng, B. et al. Parity–time-symmetric whispering-gallery microcavities. *Nature Phys.* **10**, 394–398 (2014).

[42] Miri, M.-A. & Alù, A. Exceptional points in optics and photonics. *Science* **363**, eaar7709 (2019).

[43] Li, A. *et al*. Exceptional points and non-Hermitian photonics at the nanoscale. *Nat. Nanotechnol.* **18**, 706–720 (2023).

[44] Peng, B. et al. Loss-induced suppression and revival of lasing. *Science* **346**, 328–332 (2014).

[45] Wiersig, J. Review of exceptional point-based sensors. *Photon. Res.* **8**, 1457 (2020).

[46] Mao, W., Fu, Z., Li, Y., Li, F. & Yang, L. Exceptional–point–enhanced phase sensing. *Sci. adv.* **10**, eadl5037 (2024).

[47] Wang, C., Sweeney, W. R., Stone, A. D. & Yang, L. Coherent perfect absorption at an exceptional point. *Science* **373**, 1261–1265 (2021).

[48] Zhang, B. et al. Dissipative Kerr single soliton generation with extremely high probability via spectral mode depletion. *Front. Optoelectron.* **15**, 48 (2022).

[49] Komagata, K. et al. Dissipative Kerr solitons in a photonic dimer on both sides of exceptional point. *Commun. Phys.* **4**, 159 (2021).

[50] Kim, C. et al. Parity-time symmetry enabled ultra-efficient nonlinear optical signal processing. *eLight* **4**, 6 (2024).

[51] Vernon, Z. et al. Truly unentangled photon pairs without spectral filtering. *Opt. Lett.* **42**, 3638 (2017).

[52] Liu, Y. et al. High-spectral-purity photon generation from a dual-interferometer-coupled silicon microring. *Opt. Lett.* **45**, 73 (2020).

[53] Tison, C. C. et al. Path to increasing the coincidence efficiency of integrated resonant photon sources. *Opt. Express* **25**, 33088 (2017).

[54] Lu, X., Jiang, W. C., Zhang, J. & Lin, Q. Biphoton Statistics of Quantum Light Generated on a Silicon Chip. *ACS Photonics* **3**, 1626–1636 (2016).

[55] Jiang, M.-H. et al. Quantum storage of entangled photons at telecom wavelengths in a crystal. *Nat. Commun.* **14**, 6995 (2023).

[56] Thew, R. T., Acín, A., Zbinden, H. & Gisin, N. Bell-Type Test of Energy-Time Entangled Qutrits. *Phys. Rev. Lett.* **93**, 010503 (2004).

[57] Brown, R. H. & Twiss, R. Q. LXXIV. A new type of interferometer for use in radio astronomy. *The London, Edinburgh, and Dublin Philosophical Magazine and Journal of Science* **45**, 663–682 (1954).

[58] Hu, X.-M., Guo, Y., Liu, B.-H., Li, C.-F. & Guo, G.-C. Progress in quantum teleportation. *Nat. Rev. Phys.* **5**, 339–353 (2023).

[59] Sibson, P. et al. Chip-based quantum key distribution. *Nat. Commun.* **8**, 13984 (2017).

[60] Llewellyn, D. et al. Chip-to-chip quantum teleportation and multi-photon entanglement in silicon. *Nat. Phys.* **16**, 148–153 (2020).

[61] Lu, H.-H., Liscidini, M., Gaeta, A. L., Weiner, A. M. & Lukens, J. M. Frequency-bin photonic quantum information. *Optica* **10**, 1655 (2023).

[62] Spring, J. B. et al. Boson Sampling on a Photonic Chip. *Science* **339**, 798–801 (2013).

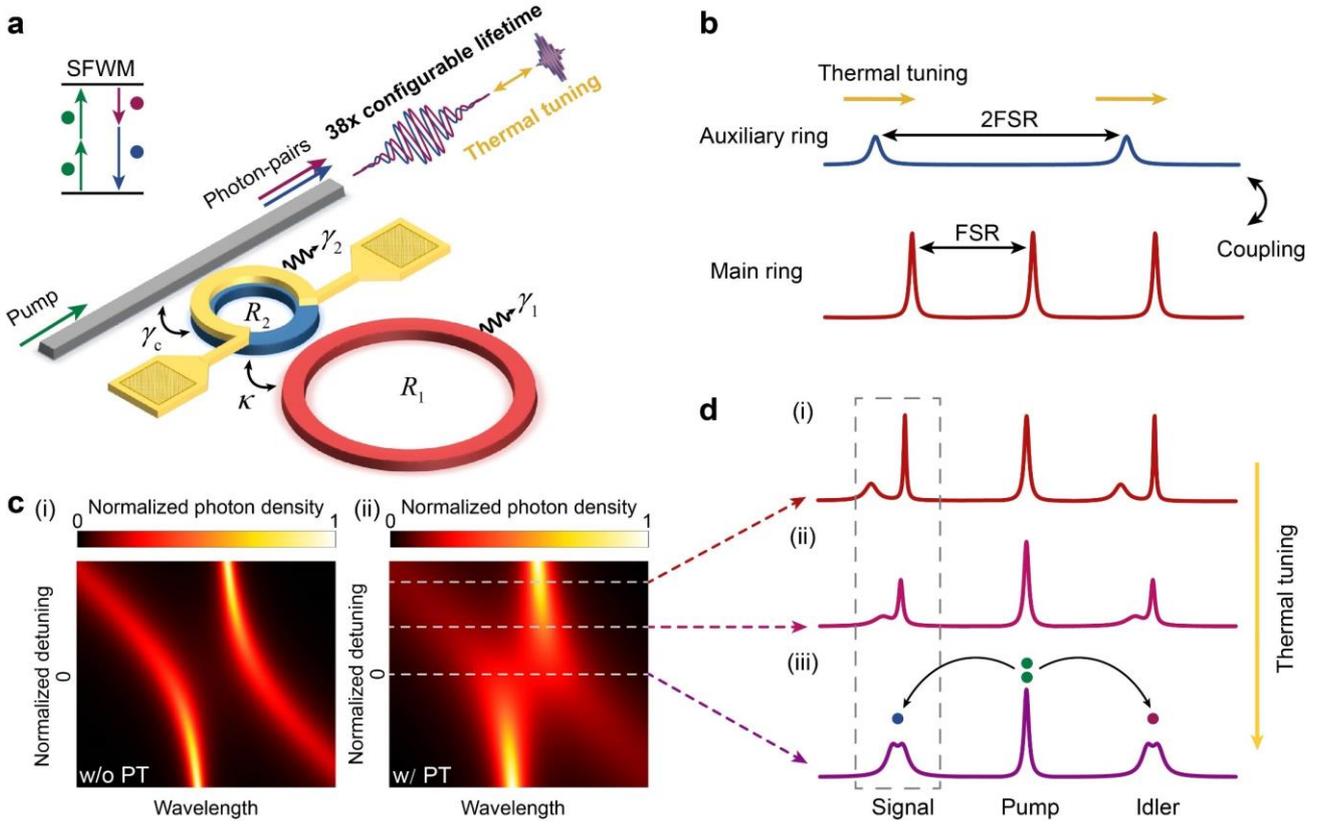

**Fig. 1. Coupled PT-symmetric microresonators system and the concept of configurable photon lifetime. a** Sketch of the coupled dual-ring system. The main ring (red) with a radius of $R_1$ is coupled to the auxiliary ring (blue) with a radius of $R_2$. The intrinsic decay of the main and auxiliary ring is $\gamma_1$ and $\gamma_2$, respectively. The coupling strength between two rings is $\kappa$. The thermal tuning module is fabricated on top of the auxiliary ring. The auxiliary ring is coupled with the bus waveguide with the coupling decay rate $\gamma_c$. Photon pairs are generated via the spontaneous four wave-mixing (SFWM) that occurs in the main ring. A 38-fold configurable photon lifetime is achieved by controlling the detuning between the two rings via the thermal tuning module. **b** Normalized intracavity photon density of state (DOS) of the main ring, with yellow arrows indicating the thermal tuning of the resonance of the auxiliary ring (redshift). **c** Photon DOS evolutions (i) without (w/o) and (ii) with (w/) PT-symmetry. **d** Detailed photon DOS at different detuning statuses marked by the dashed lines in **c** (ii).

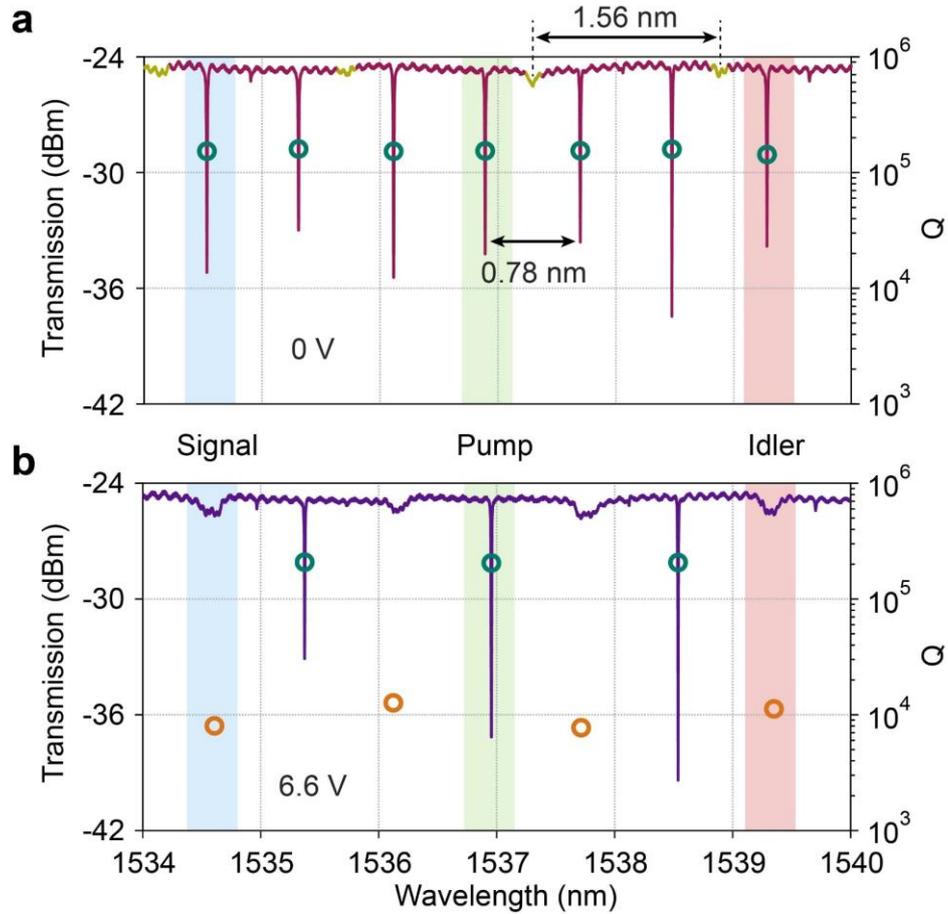

**Fig. 2. Spectra characterizations. a** System transmission spectrum under the 0 V bias (left axis). A series of high-Q resonances come from the main ring, and the highlighted yellow lines are the resonances of the auxiliary ring. **b** System transmission spectrum under the 6.6 V bias (left axis). The green circles in **a** and **b** are the marks of the high Q-factors of the main ring's resonances (right axis). The orange circles in **b** are the low Q-factors of merged resonances. The dashed blue, green, and red regions represent the locations of signal, pump, and idler resonances.

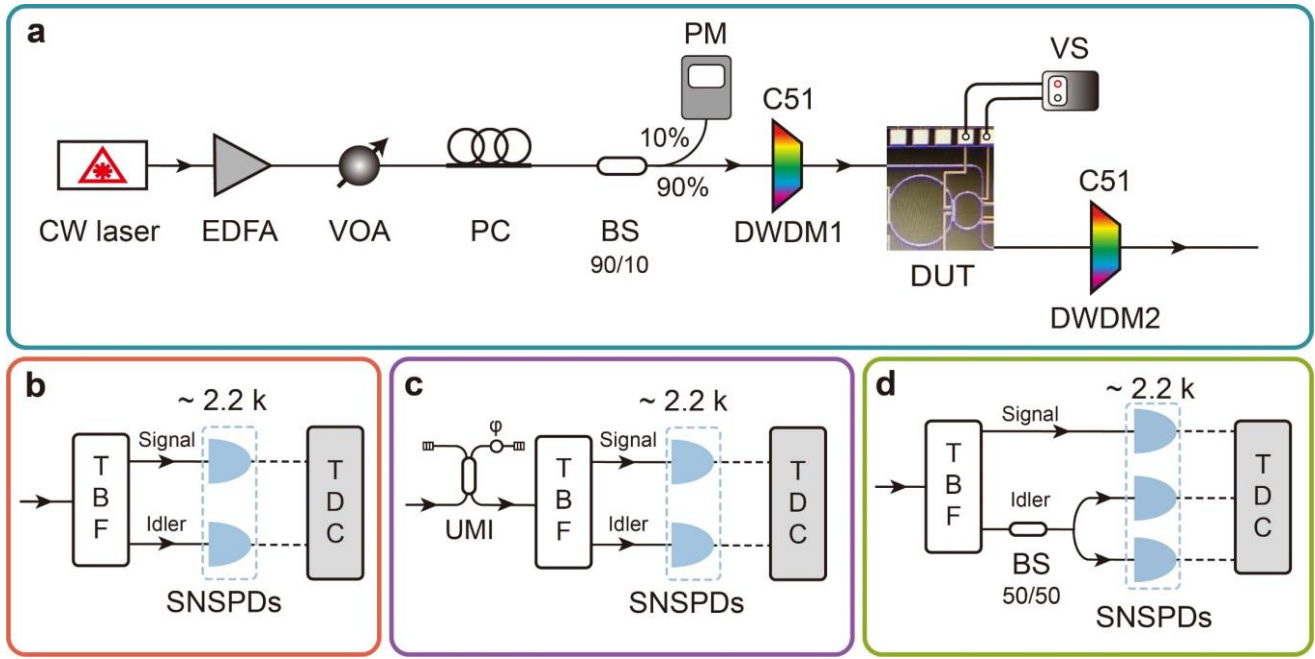

**Fig. 3. Experimental setups. a** Generation of correlated photon pairs from PT symmetry-enabled lifetime tunable device. **b** Test of correlation properties (coincidence counts). **c** Characterization of energy-time entanglement with two-photon interference. **d** Test of heralded single-photon using Hanbury Brown-Twiss (HBT) interferometer. CW: continuous wave, EDFA: erbium-doped fiber amplifier, VOA: variable optical attenuator, PC: polarization controller, BS: beam splitter, PM: power meter, DWDM: dense wavelength division multiplexer, DUT: device under test, VS: voltage source, TBF: tunable bandwidth filter, SNSPDs: superconducting nanowire single-photon detectors, TDC: time-to-digital converter, UMI: unbalanced Michaelson interferometer.

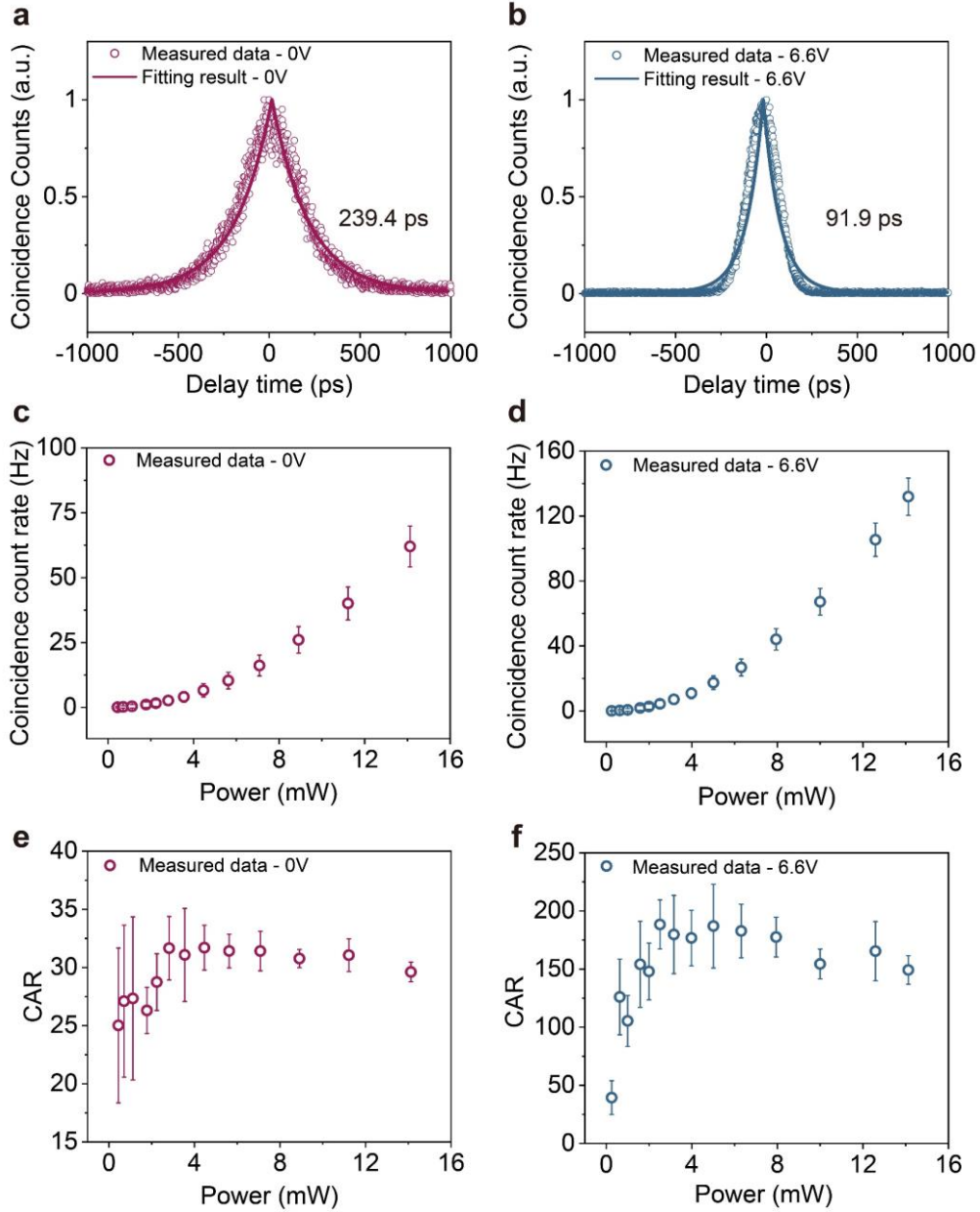

**Fig. 4. Experimental results of photon-pairs generation and lifetime adjustment. a** Normalized coincidence counts without thermal tuning (0 V bias). The photons generated from high-Q resonances are fitted with the 1/e coherent time of $\tau_{1/e} = 239.4$ ps. **b** Normalized coincidence counts with thermal tuning (6.6 V bias). The fitted 1/e coherent time is $\tau_{1/e} = 91.9$ ps. **c** and **d** are the coincidence count rates with respect to the on-chip pump powers without and with thermal tuning, respectively. **e** and **f** are the corresponding CARs.

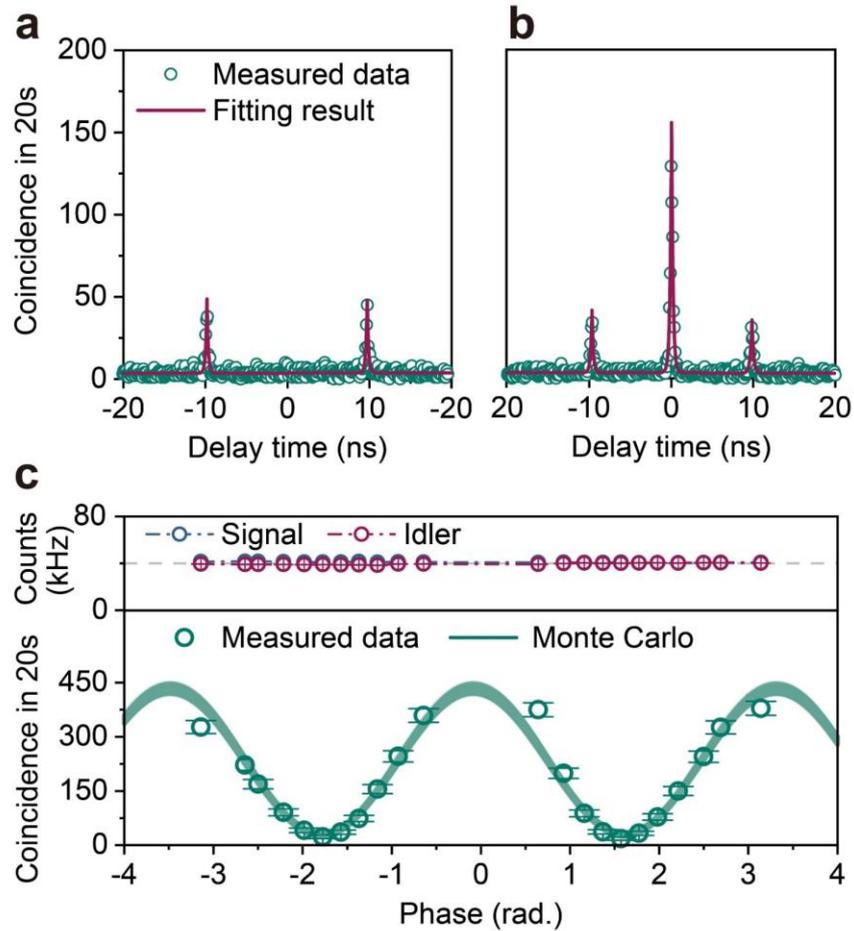

**Fig. 5. Short-lived photon pairs test of two-photon interference and energy-time entanglement.**
**a** Two-photon destructive interference with respect to delay time. **b** Two-photon constructive interference with respect to delay time. **c** Single side counts of signal and idler photons (upper panel), as well as the two-photon interference fringe with a 1000-time Monte Carlo method fit (lower panel).

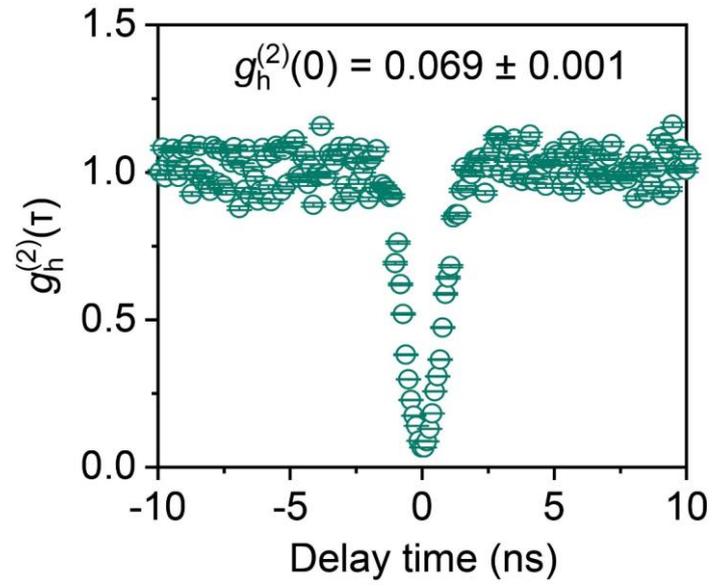

**Fig. 6. Measurement of heralded second-order autocorrelation**. $g_h^{(2)}(0)$ is tested to be $0.0069 \pm 0.001$, showing good single photon properties.